\documentclass[twocolumn]{book}
\usepackage[dvips]{graphicx,color}
\usepackage{makeidx,universe}

\newcommand{\met}{\mbox{\ensuremath{\slash\kern-.7emE_{T}}}}

\makeauthorindex

\BookTitle{Proceedings of the XXIX PHYSICS IN COLLISION}

\CopyRight{\copyright 2009 by Universal Academy Press, Inc.}

\begin{document} 

\pagenumbering{arabic}

\chapter{%
{\LARGE \sf
Searches for New Physics with High Energy Colliders } \\
{\normalsize \bf 
Emmanuel Sauvan$^{1}$} \\
{\small \it \vspace{-.5\baselineskip}
(1) CPPM, CNRS/IN2P3 et Aix-Marseille Universit\'e, 163 Av. de Luminy, F-13288 Marseille, France 
}
}


\AuthorContents{E. Sauvan}

\AuthorIndex{Sauvan}{E.}

  \baselineskip=10pt 
  \parindent=10pt    

\section*{Abstract} 

Recent experimental results of searches for new phenomena performed at high energy colliders are reviewed.
The results reported are based on data samples of up to $1$ fb$^{-1}$ and $4$ fb$^{-1}$ collected at HERA and at the Tevatron, respectively.
No significant evidence for physics beyond the Standard Model has been found and limits at the $95$\% confidence level have been set on the mass and couplings of several possible new particles.

\section{Introduction} 

For more than a century, high-energy collisions of particles have been a golden method of investigating the structure of matter. Together with precision studies of heavy meson decays, primarily at lower energy colliders, these experiments have lead to the consolidation of the Standard Model (SM) of strong, weak and electroweak interactions.
Although remarkably confirmed by present experimental results, the SM remains unsatisfactory and incomplete.
Many questions are unexplained by the SM.
For example, the SM does not explain the quantisation of the electromagnetic charge, or the observed replication of the three fermion families.
It explains neither the origin of fermion masses nor the observed hierarchy between them.
No candidate for the dark matter exists in the SM.
%
%
Many models of new physics have been proposed to address these issues, the most popular among them being the supersymmetry.

The corrections to the Higgs mass indicate that new physics may be at an energy scale of the order of $1$~TeV. It would therefore be possible to see effects beyond the SM at present high-energy colliders.
They indeed provide high sensitivity to new phenomena, allowing new massive particles to be directly produced or the effect of new interactions interfering with SM processes to be studied. 

At HERA (Hadron Electron Ring Anlage) electrons (or positrons) collide with protons at a centre-of-mass energy of $\sqrt s \simeq 320$~GeV.
During the two running periods of HERA from $1994$ to $2000$ and from $2003$ to $2007$, respectively, the H1 and ZEUS experiments have each recorded $\sim 0.5$ fb$^{-1}$ of data in total, shared between $e^+p$ and $e^-p$ collision modes.
These high energy electron-proton interactions provide a testing ground for the SM, complementary to $e^+e^-$ and $p\bar{p}$ scattering studied at the LEP and at the Tevatron, respectively.

The Tevatron proton-antiproton collider delivers data to the CDF and D\O\ experiments with a collision energy in the centre-of-mass of $1.96$~TeV. Since the start of its second running phase, a total integrated luminosity of \mbox{$\sim 6$~fb$^{-1}$} has been collected by each experiment and up to \mbox{$\sim 4$~fb$^{-1}$} are presently used to search for new phenomena.
Before the start of the Large Hadron Collider (LHC), the Tevatron is the only high-energy collider taking data, with steadily increasing integrated luminosities.

\section{Model-Driven Searches}

\subsection{Compositness}

The observed replication of three fermion families motivates the possibility of a new scale of matter yet unobserved.

In the line of historical scattering experiments which lead to the successive discoveries of the different substructures of matter, the nucleus, nucleons and quarks, scattering of point-like electrons on quarks at HERA can be used to search for possible substructure of quarks. 
A finite charge radius for the quark would modify the $d\sigma/dQ^2$ neutral current (NC) deep-inelastic scattering (DIS) cross section of the electron on a quark  with a form factor type term $(1-R^2_q/6 \; Q^2)^2$, where $R_q$ is the root-mean-square radius of the electroweak charge of the quark.
At the highest $Q^2$ values, no deviations between the data and the SM prediction, derived from the DGLAP evolution of parton density functions determined at lower $Q^2$, was observed by the H1 and ZEUS experiments, using their full data set~\cite{H1_CI,ZEUS_CI}.
The most stringent constraint on the the quark radius of $R_q < 0.63 \; 10^{-18}$~m is derived by the ZEUS experiment~\cite{ZEUS_CI}.

\begin{figure}[!htbp]
  \begin{center}
  \includegraphics[width=0.4\textwidth]{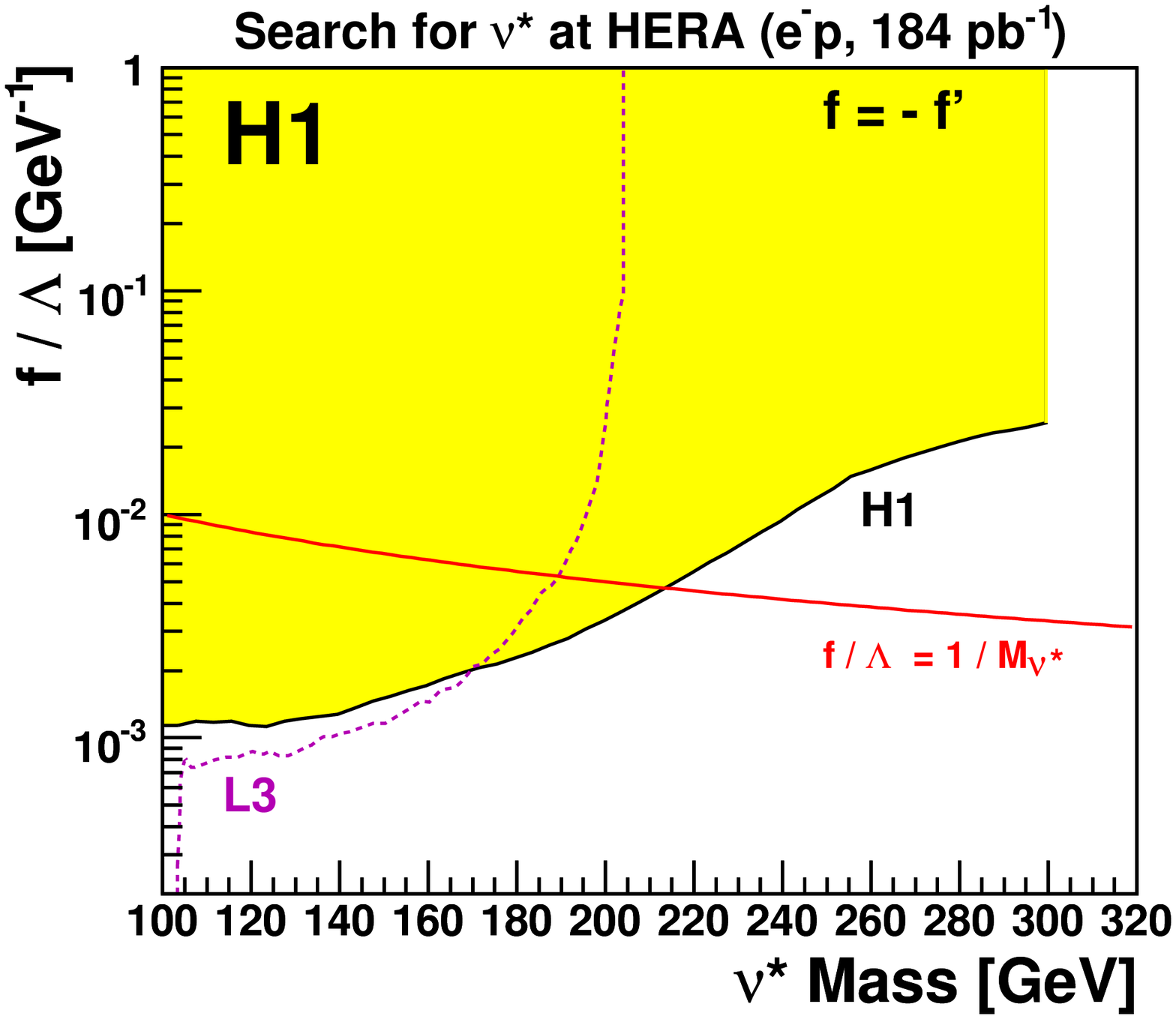}\put(-25,35){{ (a)}}\\
  \includegraphics[width=0.4\textwidth]{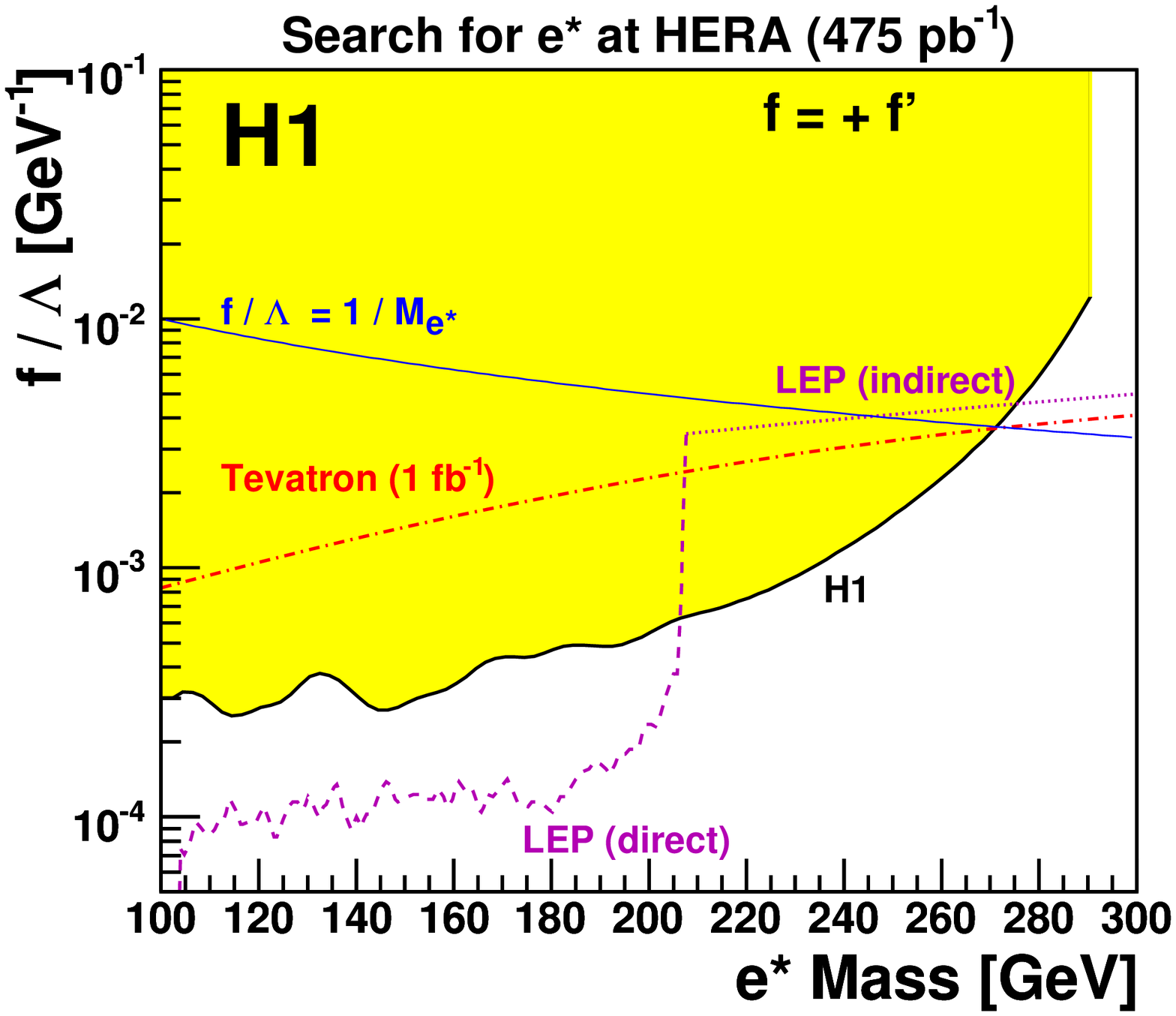}\put(-25,35){{ (b)}}\\
  \includegraphics[width=0.4\textwidth]{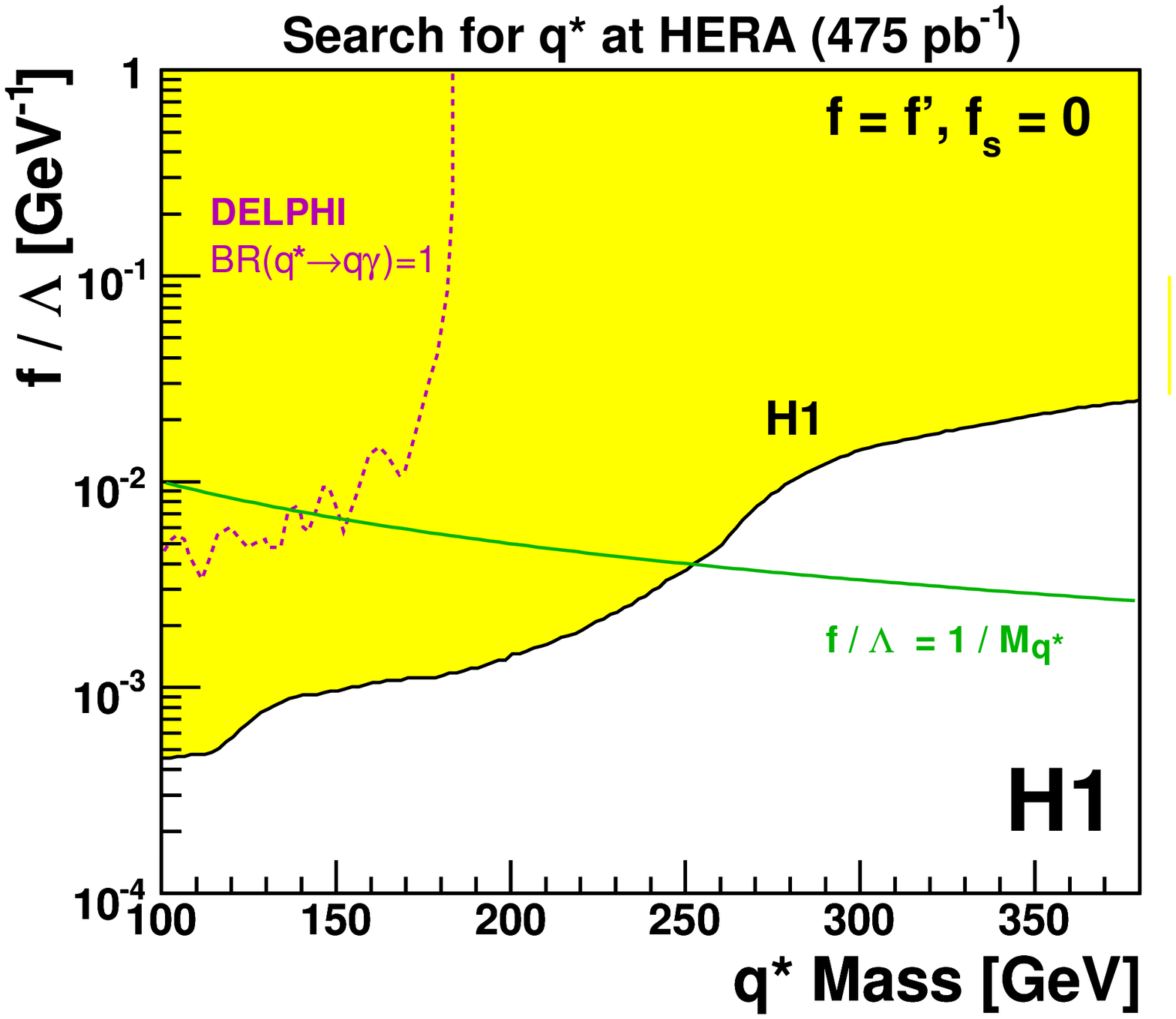}\put(-100,35){{ (c)}}
  \end{center}
  \vspace{-1pc}
  \caption{Exclusion limits on the coupling $f/\Lambda$ as a function of the mass of the excited neutrino (a), electron (b) and quarks (c). 
The new limits set by H1 are represented by the shaded area.
Values of the couplings above the curves are excluded.}
  \label{fig:exc_fermions}
\end{figure}

An unambiguous signature for a new scale of matter would be the direct observation of excited states of fermions ($f^*$), via their decay into a gauge boson and a fermion. Effective models describing the interaction of excited fermions with standard matter have been proposed~\cite{Hagiwara:1985wt,Boudjema:1992em,Baur:1989kv}.
In the models~\cite{Hagiwara:1985wt,Boudjema:1992em} the interaction of an $f^*$ with a gauge boson is described by a magnetic coupling proportional to $1/\Lambda$ where $\Lambda$ is a new scale. Proportionality constants $f$, $f'$ and $f_s$ result in different couplings to $U(1)$, $SU(2)$ and $SU(3)$ gauge bosons.

The H1 experiment has carried out searches for excited neutrinos ($\nu^*$), electrons ($e^*$) and quarks ($q^*$) using its full data set~\cite{nustar_H1,estar_H1,qstar_H1}. The total luminosity analysed amount to up to $475$ pb$^{-1}$. The new bounds on the $\nu^*$ and $e^*$ masses obtained as a function of $f/\Lambda$ are presented in Figure~\ref{fig:exc_fermions}(a) and (b), under the assumptions $f = - f'$ and $f = + f'$, respectively. 
Assuming $f/\Lambda = 1/M_{\nu^*}$ and $f = - f'$, masses below $213$~GeV are ruled out for $\nu^*$. Excited electrons of mass below $272$~GeV are excluded if we assume $f/\Lambda = 1/M_{e^*}$ and $f = + f'$.
Searches for $q^*$ performed at HERA are complementary to Tevatron results, since at a $p\bar{p}$ collider excited quarks are dominantly produced in a quark-gluon fusion mechanism, which requires $f_s \neq 0$.
For $f/\Lambda = 1/M_{q^*}$ and $f=f'$ and $f_s =0$ excited quarks with a mass below $252$~GeV are excluded by H1.
As observed in Figure~\ref{fig:exc_fermions}, the H1 analysis has probed new parameter space regions, and the limits set extend previous bounds reached at LEP and Tevatron colliders.

\subsection{Leptoquarks}

An intriguing characteristic of the Standard Model is the observed symmetry between the lepton and quark sectors.
New symmetries connecting the two sectors are therefore introduced in various unifying theories beyond the SM, leading to the appearance of leptoquarks (LQs). 
LQs are new scalar or vector color-triplet bosons, carrying a fractional electromagnetic charge and both a baryon and a lepton number. 
Several types of LQs might exist, differing in their quantum numbers.
A classification of LQs has been proposed by Buchm\"uller, R\"uckl and Wyler (BRW)~\cite{brw_lq} under the assumption that LQs have pure chiral couplings to SM fermions of a given family. The interaction of the LQ with a lepton-quark pair is of Yukawa or vector nature and is parametrised by a coupling $\lambda$.
Leptoquarks decay into a quark and a charged or neutral lepton, with branching fractions $\beta$ and $(1-\beta)$, respectively.

\begin{figure}[htbp]
  \begin{center}
  \includegraphics[width=0.4\textwidth]{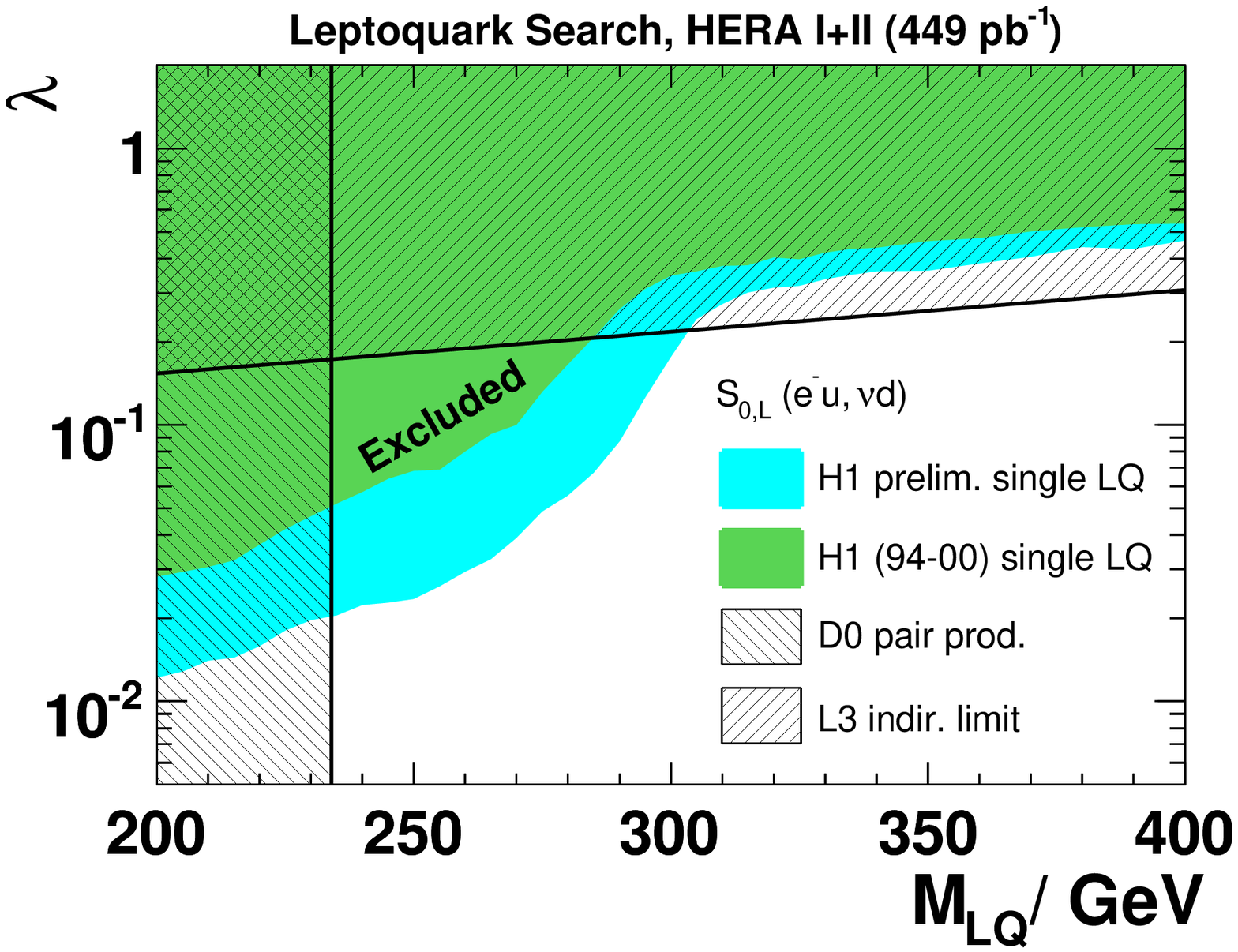}\put(-38,35){{(a)}}\\
    \includegraphics[width=0.4\textwidth]{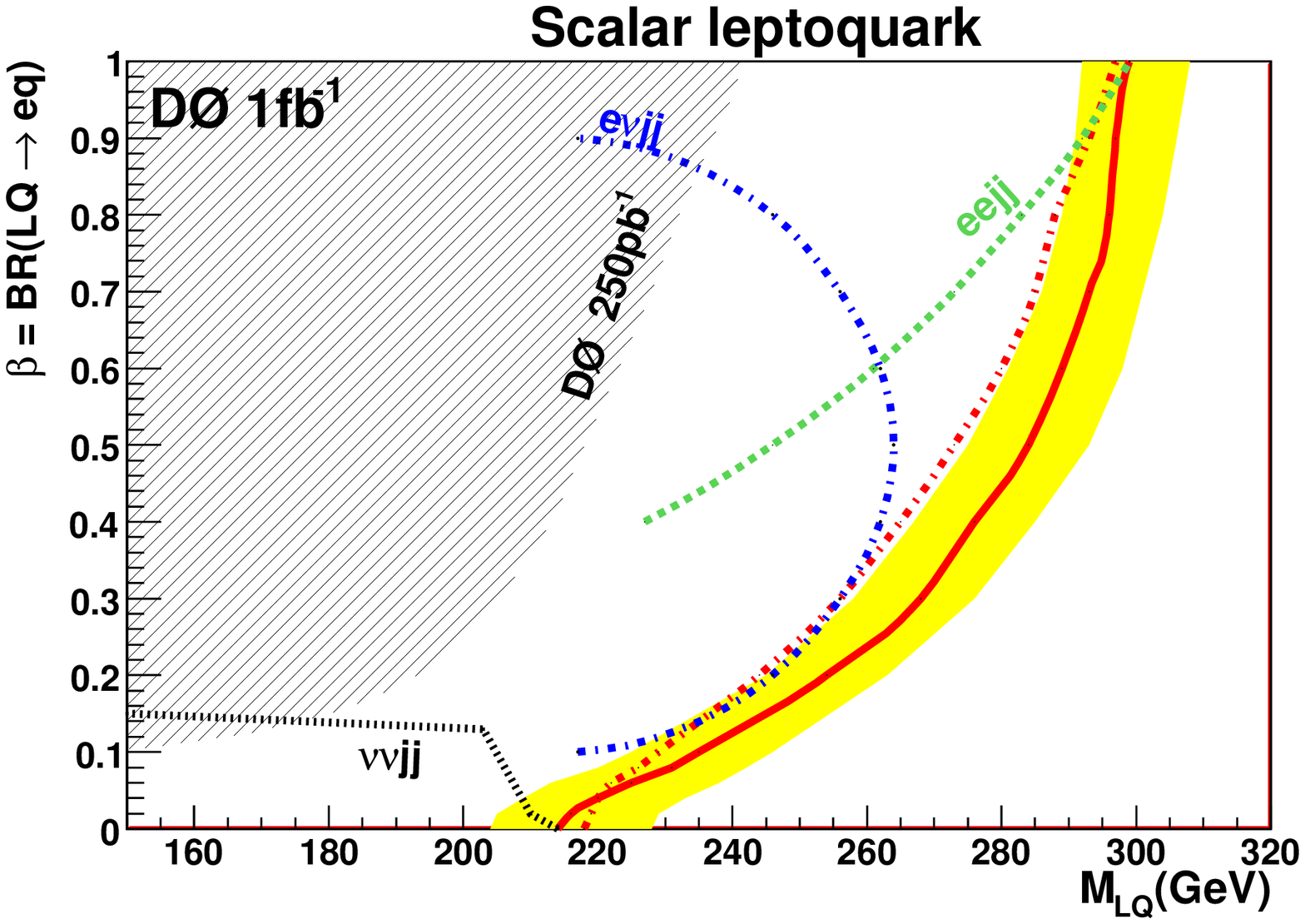}\put(-40,25){{(b)}}
  \end{center}
  \vspace{-1pc}
  \caption{(a) Exclusion limit set by H1 on the coupling $\lambda$ as a function of the mass for $S_{0,L}$ leptoquarks.(b) Limits on the mass of first generation scalar LQs as a function of $\beta$ obtained by D\O~\cite{LQ1_D0}. The observed limit is represented by the full line and the expected limit by the dot-dashed line.}
  \label{fig:D0_LQ1}
\end{figure}

In $ep$ collisions at HERA, first generation LQs might be singly produced from the fusion of the incoming lepton with a quark from the proton. 
LQs might therefore be observed as a resonant peak in the lepton-jet mass distribution of neutral or charge current DIS events.
No such signal has been observed by the H1 experiment in a search using up to $449$~pb$^{-1}$ of data and limits on the LQ mass depending on the $\lambda$ coupling have been set~\cite{H1_LQs} (see figure~\ref{fig:D0_LQ1}(a)).

At the Tevatron, LQs might be pair produced in $q\bar{q}$ interactions via their coupling to gluons.
Depending on their decay, the possible final states are $\ell^+ \ell^- q\bar{q}$, $\ell \nu q\bar{q}$ or $\nu\bar{\nu}q\bar{q}$ with respective branching fractions $\beta^2$, $2\beta(1-\beta)$ or $(1-\beta)^2$.
Searches have been performed by the D\O\ Collaboration for LQs of all three generations.
For the first generation, the search was performed in the $eejj$ and $e\nu jj$ channels and LQs with a mass below $299$~GeV have been excluded for $\beta =1$~\cite{LQ1_D0}.
The limit on the LQ mass as a function of $\beta$ is shown in figure~\ref{fig:D0_LQ1}(b).
Lowest values of $\beta$ are probed in a search for two acoplanar jets and \mbox{\ensuremath{\slash\kern-.7emE_{T}}}~\cite{Abazov:2008at}.
Leptoquarks decaying with a large branching ratio in $\nu q$ are not easily probed at the Tevatron due to large background. In such cases, if $\lambda$ is reasonably large, HERA experiments can still provide a better sensitivity.

A search for second generation LQs has also been performed by D\O\ in the $\mu \mu jj$ and $\mu \nu jj$ channels~\cite{Abazov:2008np}. The mass limit obtained for $\beta =1$ is $316$~GeV.
Third generation LQs were searched for in the $\tau \tau bb$ channel, with one $\tau$ decaying to a muon and the other to hadrons. The mass limit obtained for $\beta =1$ is $210$~GeV~\cite{Abazov:2008jp}.
Third generation LQs are also searched via their decay into $\nu \bar{\nu} b \bar{b}$ in acoplanar jet topologies, using heavy flavour tagging. A mass limit of $252$~GeV for $\beta=0$ is reached in this case by D\O\ using $4$~fb$^{-1}$ of data~\cite{D0_LQ3_bb}.

\subsection{Sypersymmetry}

Supersymmetry (SUSY) is one of the most attractive extension of the SM.
It allows for the unification of the four known forces at the GUT scale, it is the only non-trivial extension of the Lorentz-Poincar\'e group and provides a possible candidate for the dark matter, the lightest supersymmetric particle (LSP).
In SUSY, every SM particle has a superpartner differing in spin by $1/2$. 
The SUSY partners are assigned an $R$-parity $R_P =  (-1)^{(3B+L+2S)} = -1$, in contrast to the SM particles of $R_P = +1$.
In most cases $R$-parity is assumed to be conserved and sparticles are therefore produced in pair and decay to SM particles and to the LSP. 
Since the LSP is stable and weakly interacting, the final state exhibits missing energy.

Since superpartners have not been yet observed, SUSY cannot be an exact symmetry and different breaking mechanisms have been proposed. Under certain assumptions they allow to reduce the large number ($> 100$) of free parameters in the theory to a manageable amount. 
The most widely studied model is minimal supergravity, mSUGRA. Only five parameters are needed to fully specify the model: a common scalar mass $m_0$, a common gaugino mass $m_{1/2}$, a common trilinear coupling $A_0$, the ratio $\beta$ of the vacuum expectation values of the two Higgs doublets and the sign of $\mu$, the supersymmetric mass term.

\subsubsection{Searches for squarks and gluinos}

At the Tevatron, squarks and gluinos are expected to be copiously produced via the strong interaction.

\begin{figure}[!htbp]
  \begin{center}
    \includegraphics[width=0.35\textwidth]{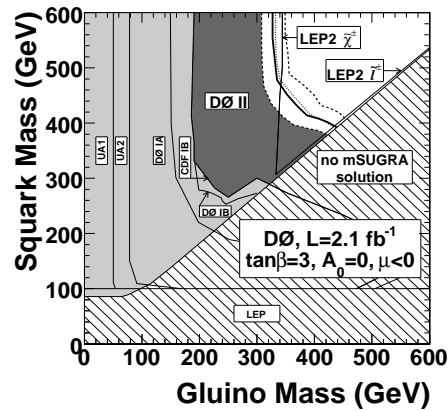}
  \end{center}
  \vspace{-1pc}
  \caption{Limits on the squark and gluino masses obtained by D\O~\cite{Abazov:2007ww} and by earlier experiments.
  The thick line corresponds to the limit obtained with nominal choices of renormalisation and factorisation scales and of parton distribution functions. The band delimited by the two dashed lines represents the uncertainty associated to different choices of scales and of PDFs.}
  \label{fig:squarks_gluinos}
\end{figure}

Due to the large production cross section, the inclusive production of squarks and gluinos is one of the most promising discovery channels for SUSY at the Tevatron.
The cascade decays of the produced squark and gluino give rise to final states with two to four jets and \mbox{\ensuremath{\slash\kern-.7emE_{T}}}, depending whether the squark mass is smaller, equal or greater than the gluino mass.
The CDF and D\O\ Collaborations have performed searches optimised for the three cases and using $\sim 2$~fb$^{-1}$ of data~\cite{Aaltonen:2008rv,Abazov:2007ww}.
As no particular excess is observed, limits are set in the two-dimensional plane ($m_{\tilde{q}}$, $m_{\tilde{g}}$) in the context of the mSUGRA model (see figure~\ref{fig:squarks_gluinos}).
Gluinos with a mass below $308$~GeV are now excluded for all squark masses and squarks lighter than $380$~GeV are excluded for all gluinos masses

\begin{figure}[htbp]
  \begin{center}
    \includegraphics[width=0.4\textwidth]{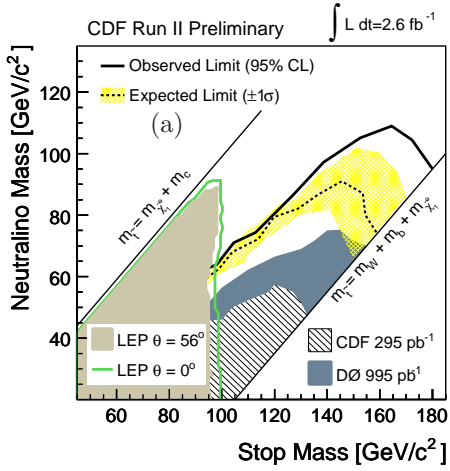}\put(-145,130){{ (a)}}\\
    \includegraphics[width=0.38\textwidth]{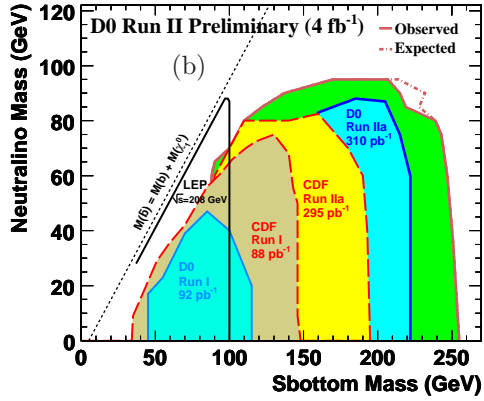}\put(-130,125){{ (b)}}
  \end{center}
  \vspace{-1pc}
  \caption{Exclusion limits on the stop (a) and sbottom (b) masses as a function of the neutralino mass obtained by CDF and D\O, respectively.}
  \label{fig_stop_sbottom}
\end{figure}

In the case of the third generation squarks, light mass eigenstates $\tilde{t}_1$ and $\tilde{b}_1$ are possible, in case of a large mixing between the right-handed and left-handed weak eigenstates.
Sbottom and stop may be produced in pair via $q\bar{q}$ annihilation or gluon-gluon fusion and the production cross section mainly depends on their mass.
They subsequently decay into a quark and the LSP, $p\bar{p} \rightarrow \tilde{t}_1 \bar{\tilde{t}}_1 \rightarrow c \tilde{\chi}_1^0 \, \bar{c} \tilde{\chi}_1^0$ and $p\bar{p} \rightarrow \tilde{b}_1 \bar{\tilde{b}}_1 \rightarrow b \tilde{\chi}_1^0 \, \bar{b} \tilde{\chi}_1^0$, leading to final states with two acoplanar charm- or $b$-jets with high \mbox{\ensuremath{\slash\kern-.7emE_{T}}}.
In a recent stop search in this channel, the CDF Collaboration exploits charm tagging of jets to reduce the background from $b$-jets~\cite{CDF_stop_ccjets}. The sensitivity achieved with $2.6$~fb$^{-1}$ of data and the absence of signal allows to exclude $\tilde{t}_1$ masses up to $180$~GeV.
The exclusion limit obtained on the stop mass depending on the neutralino mass is presented in figure~\ref{fig_stop_sbottom}(a). It extends beyond the LEP reach.
No signal of sbottom production was also observed~\cite{D0_LQ3_bb,CDF_sbottom}. The exclusion area for the sbottom obtained by the D\O\ Collaboration is shown in figure~\ref{fig_stop_sbottom}(b). It is the most sensitive such analysis to date.

\subsubsection{Searches for charginos and neutralinos}

\begin{figure}[htbp]
  \begin{center}
    \includegraphics[width=0.42\textwidth]{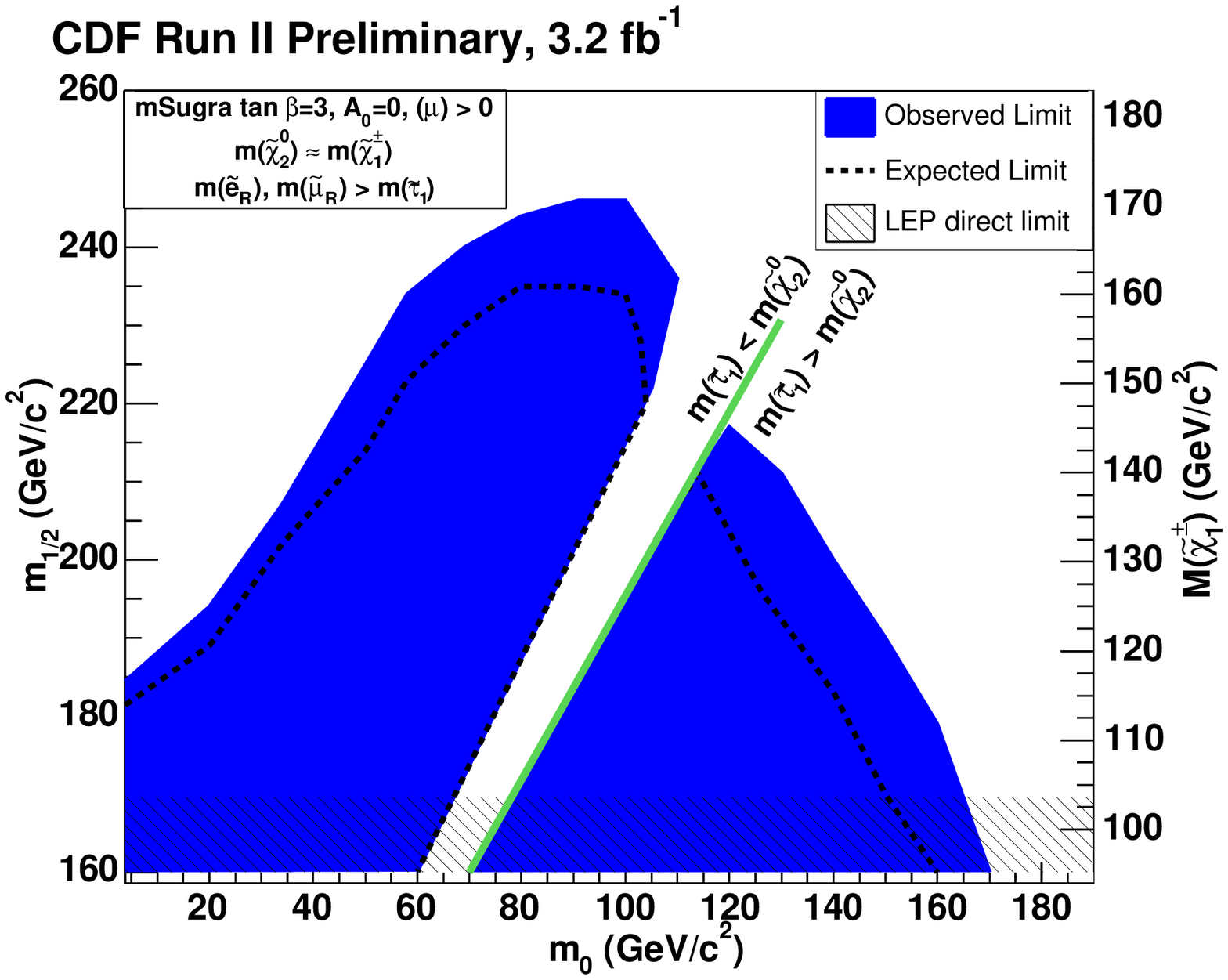}\put(-65,80){{ (a)}}\\
    \includegraphics[width=0.38\textwidth]{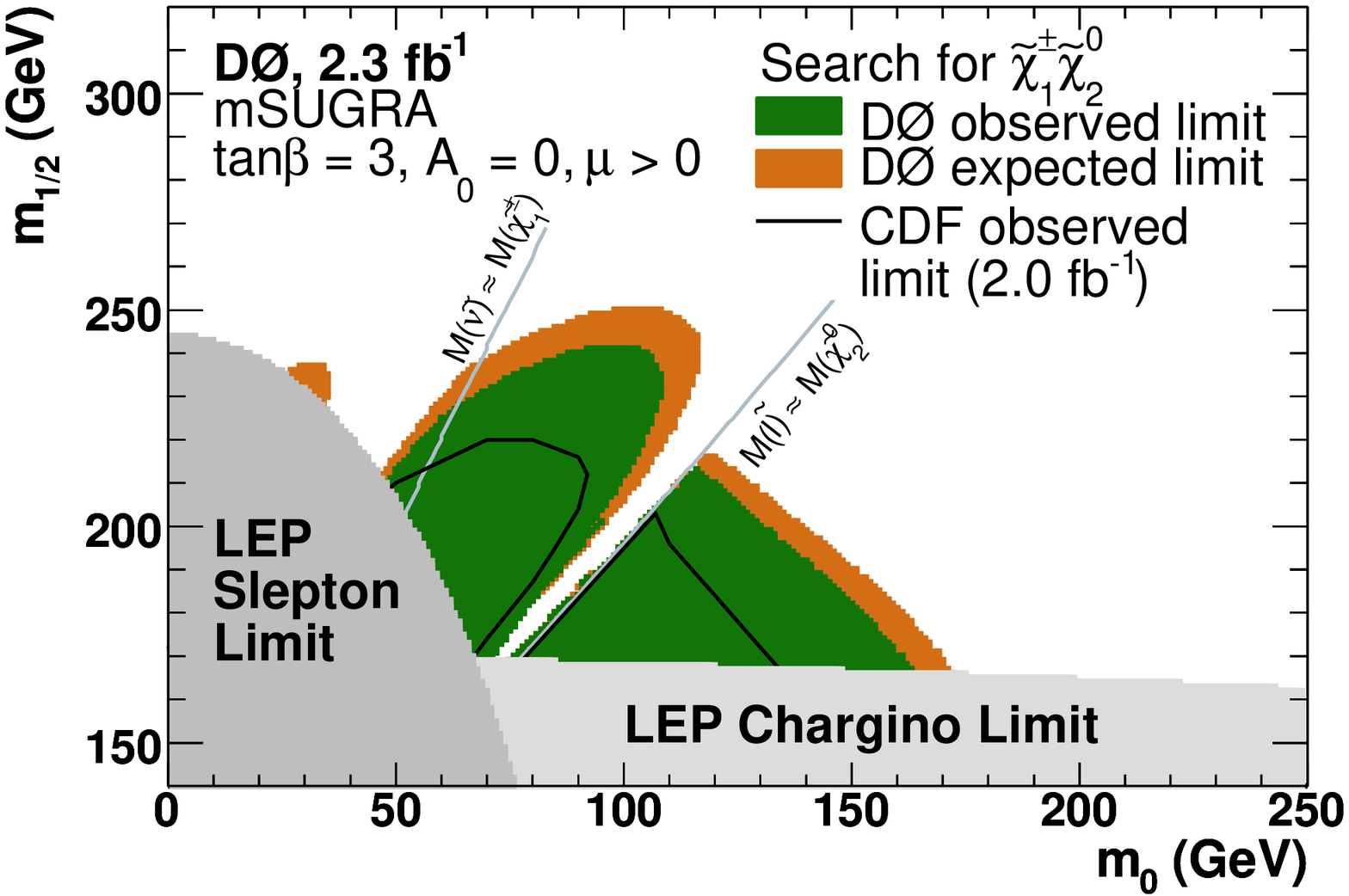}\put(-30,60){{ (b)}}
  \end{center}
  \vspace{-1pc}
  \caption{Regions in the $(m_0,m_{1/2})$ plane excluded by the CDF (a) and D\O\ (b) experiments in the trilepton search for electroweak gauginos.}
  \label{fig:gauginos}
\end{figure}

Chargino and neutralino production in $p\bar{p}$ collisions is mediated by electroweak interactions and has therefore a low production cross section.
They are best searched at the Tevatron in the associated production channel, $q\bar{q} \rightarrow W^* \rightarrow \tilde{\chi}^\pm_1 \tilde{\chi}^{0}_2 \rightarrow \ell \ell \ell \nu \tilde{\chi}^{0}_1 \tilde{\chi}^{0}_1$.  It leads to a very clean experimental signature with three leptons, of little energy, and \mbox{\ensuremath{\slash\kern-.7emE_{T}}}, in a low background environment.
Because of the small leptonic branching fractions, several final states need to be combined.
Both CDF and D\O\ experiments found no evidence for excess of events with this signature
and set limits on the mSUGRA parameters $m_0$ and $m_{1/2}$, extending the parameter space probed by LEP experiments (see figure~\ref{fig:gauginos})~\cite{Abazov:2009zi,CDF_neutralinos}. 

\subsubsection{Gauge Mediated SUSY Breaking}

\begin{figure}[htbp]
  \begin{center}
    \includegraphics[width=0.35\textwidth]{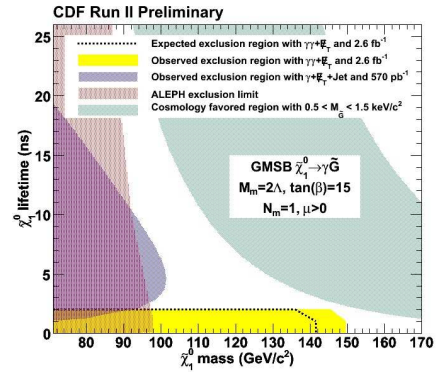}
  \end{center}
  \vspace{-1pc}
  \caption{Excluded domain in the $\tilde{\chi}_1^0$ mass and lifetime plane derived from the search for GMSB neutralino NLSP performed by CDF.
  }
  \label{fig:CDF_GMSB}
\end{figure}

In the gauge mediated SUSY breaking (GMSB) scenario, the gravitino is the LSP and the next-to-lightest SUSY particle (NLSP) may be the lightest neutralino, which decays to a gravitino and a photon, $\tilde{\chi}_1^0 \rightarrow \gamma \tilde{G}$.
At the Tevatron, pair production of SUSY particles decaying to a neutralino NLSP with negligible life time would therefore lead to final states with two acoplanar photons and \mbox{\ensuremath{\slash\kern-.7emE_{T}}}.
Searches for such events have been performed by the CDF and D\O\ Collaborations but no excess was observed over the backgrounds coming mainly from photon misidentification of from fake \mbox{\ensuremath{\slash\kern-.7emE_{T}}}~\cite{CDF_GMSB,Abazov:2007is}.
The process is then used to set a limit on the GMSB scale $\Lambda$, which also gives the scale of the gaugino masses.
For a $\tilde{\chi}_1^0$ lifetime of $0$~ns, $\tilde{\chi}_1^0$ masses below $149$~GeV are excluded by the analysis of the CDF experiment, based on $2.6$~fb$^{-1}$ of data~\cite{CDF_GMSB}.
The domain in the plane of the neutralino mass and lifetime excluded by this analysis  is compared to previous existing limits in the figure~\ref{fig:CDF_GMSB}.

\subsubsection{$R$-parity violation}

\begin{figure}[htbp]
  \begin{center}
    \includegraphics[width=0.35\textwidth]{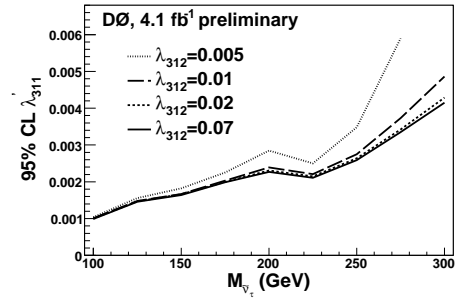}
  \end{center}
  \vspace{-1pc}
  \caption{Limits on RPV couplings as a function of the sneutrino mass obtained by D\O.
  }
  \label{fig:D0_sneutrinos}
\end{figure}

In searches discussed so far, $R$-parity was assumed to be conserved. 
If $R$-parity violation (RPV) is allowed, $\tilde{\nu}_\tau$ sneutrinos might be singly produced at the Tevatron via a non-zero $\lambda'_{311}$ Yukawa coupling to $d$-quarks, followed by  decays into two $e\mu$, $e\tau$ or $\mu\tau$ leptons.
This would produce very clean signatures with a two-lepton resonance on a low SM background environment. 
The $e\mu$ channel has been investigated by the D\O\ Collaboration using $4.1$~fb$^{-1}$ of data~\cite{D0_RPV_sneutrino} and, in the absence of a signal, limits on the mass of $\tilde{\nu}_\tau$ as a function of the Yukawa couplings $\lambda'_{311}$ and $\lambda_{312}$ have been set (see figure~\ref{fig:D0_sneutrinos}).
A similar search has been done by the CDF Collaboration, looking also at the $\mu\tau$ and $e\tau$ channels, but no deviation from the SM expectation was observed~\cite{CDF_RPV_sneutrino}.

\begin{figure}[htbp]
  \begin{center}
    \includegraphics[width=0.35\textwidth]{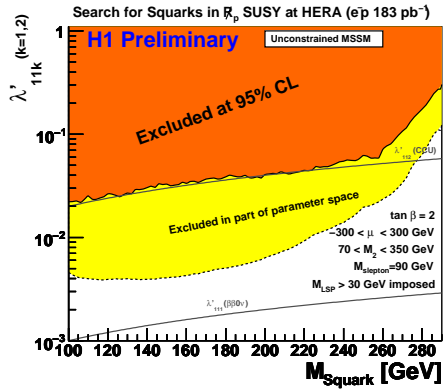}
  \end{center}
  \vspace{-1pc}
  \caption{Exclusion limits on RPV $d$-type squark production obtained by the H1 experiment.
  A scan of the SUSY parameter space has been performed. The dark shaded domain is ruled out for any value of the parameters.
  }
  \label{fig:h1_squarks}
\end{figure}

If $R$-parity is not conserved, squarks could also be resonantly produced at HERA, similarly to leptoquarks. In addition to the ``LQ-like'' decays into $eq$ and $\nu q$ the squarks also undergo cascade decays via gauginos, leading to a large number of possible final states.
A recent search for first and second generation squarks has been done by the H1 Collaboration
using its full data sample with an integrated luminosity of $438$~pb$^{-1}$~\cite{H1_RPV_squarks}.
Almost all possible final states have been searched for, taking into account direct and indirect $R$-parity violating decay modes.
In the absence of a signal, mass dependent limits on the RPV couplings $\lambda'_{1jk}$ have been derived within a phenomenological version of the minimal supersymmetric model. 
An example of limits obtained for $d$-type squarks is shown in figure~\ref{fig:h1_squarks}. 
For a Yukawa coupling $\lambda'_{1j1}$ ($\lambda'_{11k}$) of electromagnetic strength, $\lambda'_{1j1} = \sqrt{4\pi \alpha_{em}} \simeq 0.3$,   $u$-type ($d$-type) squarks up to masses $\sim 275$~GeV ($\sim 290$~GeV) are excluded.

\subsection{Extra-Dimensions}

Models based on large extra-dimensions (LED), such as the model of Arkani-Hamed, Dimopolous and Dvali (ADD)~\cite{ADD_LED} try to adress the hierarchy problem between the Planck and the electroweak scales.
In these models, gravity is allowed to propagate in a $4+n_D$ dimensional bulk space-time, while the remainder of the SM fields are confined in the $4d$ world volume.
The extra $n$ dimensions are compactified with a radius $R$. 
In the simplest model, the fundamental Planck scale $M_D$ in the $4+n_D$ space is then related to the Planck scale by $M^2_{Pl} = 8\pi R^{n_D} M_D^{n_D+2}$.
If the size of the extra-dimensions is small, $M_D$ could be of the order of $1$~TeV and its effects could be visible at present colliders.
Since it propagates in the extra-dimension, the graviton observed in $4d$ manifests itself as a tower of Kaluza-Klein (KK) excitations which form a quasi-continuum in mass.

In $p\bar{p}$ collisions, such KK gravitons can be directly produced together with a quark, a gluon or a photon  and escape into the bulk space, leading to large missing transverse energy.
The corresponding signatures are then a mono-jet, or a mono-photon, with \mbox{\ensuremath{\slash\kern-.7emE_{T}}}.
Using $2$~fb$^{-1}$ of data, CDF has recently published an analysis combining searches for mono-jet and mono-photon topologies and set limits on the effective Plank scale between $M_D > 1.4$~TeV and $M_D > 0.94$~TeV for a number of extra dimension ranging from $2$ to $6$~\cite{Aaltonen:2008hh}. 
A similar search for mono-photon events was performed by the D\O\ Collaboration in $2.7$~fb$^{-1}$ of data, leading to limits on $M_D$ of $0.97$~TeV and $0.831$~TeV for $n_D=2$ and $n_D=6$, respectively~\cite{D0_LED_direct}.

An alternative way to search for large extra-dimensions is to look for the effect of graviton exchange and its interference with SM processes.
An effective coupling of the form $\lambda/M_S^4$, with a parameter $\lambda$ which is expected to be close to one, needs to be introduced to perform calculations.
Therefore, such effects due to graviton exchanges allows to probe an effective scale $M_S$, but not the fundamental scale $M_D$. Nevertheless, this effective scale $M_S$ should not differ too much from $M_D$.
At the Tevatron, such indirect effects may induce an enhancement of the production cross section of fermion or boson pairs. This was searched for in the $ee$ and $\gamma \gamma$ final states by the D\O\ experiment~\cite{Abazov:2008as}. Combining both analyses the constraint $M_S > 1.62$~TeV is obtained, using the GRW formalism~\cite{Giudice:1998ck}.
A recent analysis from the D\O\ experiment uses high $P_T$ di-jet events and the jet angular distributions, via the variable $\chi = \exp{|y_1 - y_2|}$, where $y_1$ and $y_2$ are the rapidities of the two jets, to constraint the existence of extra-dimensions~\cite{Abazov:2009mh}. 
A lower limit of $M_S > 1.66$~TeV is obtained, which is the most stringent bound so far.

At HERA, after summing the effects of graviton excitations in the extra-dimensions, the graviton contribution could be visible as a contact interaction contributing to the $eq \rightarrow eq$ scattering.
The effect could therefore be visible in the NC DIS cross section, introducing terms of the form $\eta_G = \pm 1/ M^4_S$ in the Lagrangian. 
This was searched for by the ZEUS experiment using its full data sample~\cite{ZEUS_CI}. 
A good agreement of the measured cross section with the SM expectation is observed, allowing to set a lower limit of $M_S > 0.94$~TeV.

\section{Model-Independent Searches}

A large variety of possible extensions to the SM exists, which often predict similar experimental signatures. 
Searches for new physics presented above compare data to the predictions of these specific models.
A complementary approach which now tends to develop is followed in signature based searches. In this case differences between data and SM expectations are looked for in various event topologies. 
For a given final state, in the absence of deviations, limits can then be set on different exotic models.
As another advantage, such model independent analyses do not rely on any a priori definition of expected signatures for exotic phenomena.
Therefore, they also address the important question of whether unexpected phenomena might occur through a new pattern, not predicted by existing models.
Following this approach, signals for new phenomena can be searched for through resonances in invariant mass distributions of two SM particles, in final states corresponding to rare SM processes or in topologies containing a certain type of particle, as e.g. a photon.
More generally, a scan at high transverse momenta of all possible final states can also be performed.

\subsection{Searches for New Resonances}

\begin{figure}[htbp]
  \begin{center}
    \includegraphics[width=0.35\textwidth]{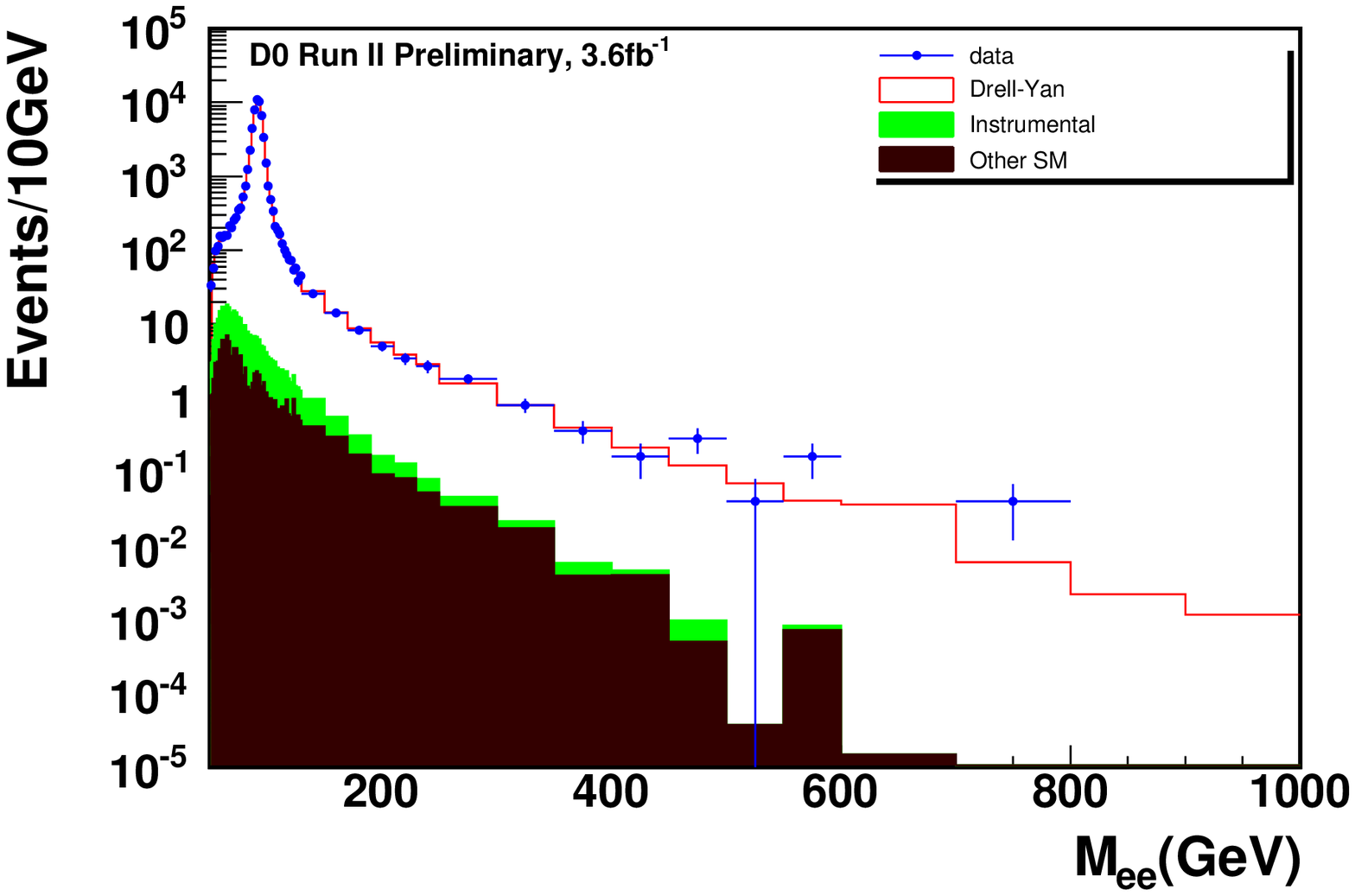}\put(-40,70){{ (a)}}\\
    \includegraphics[width=0.35\textwidth]{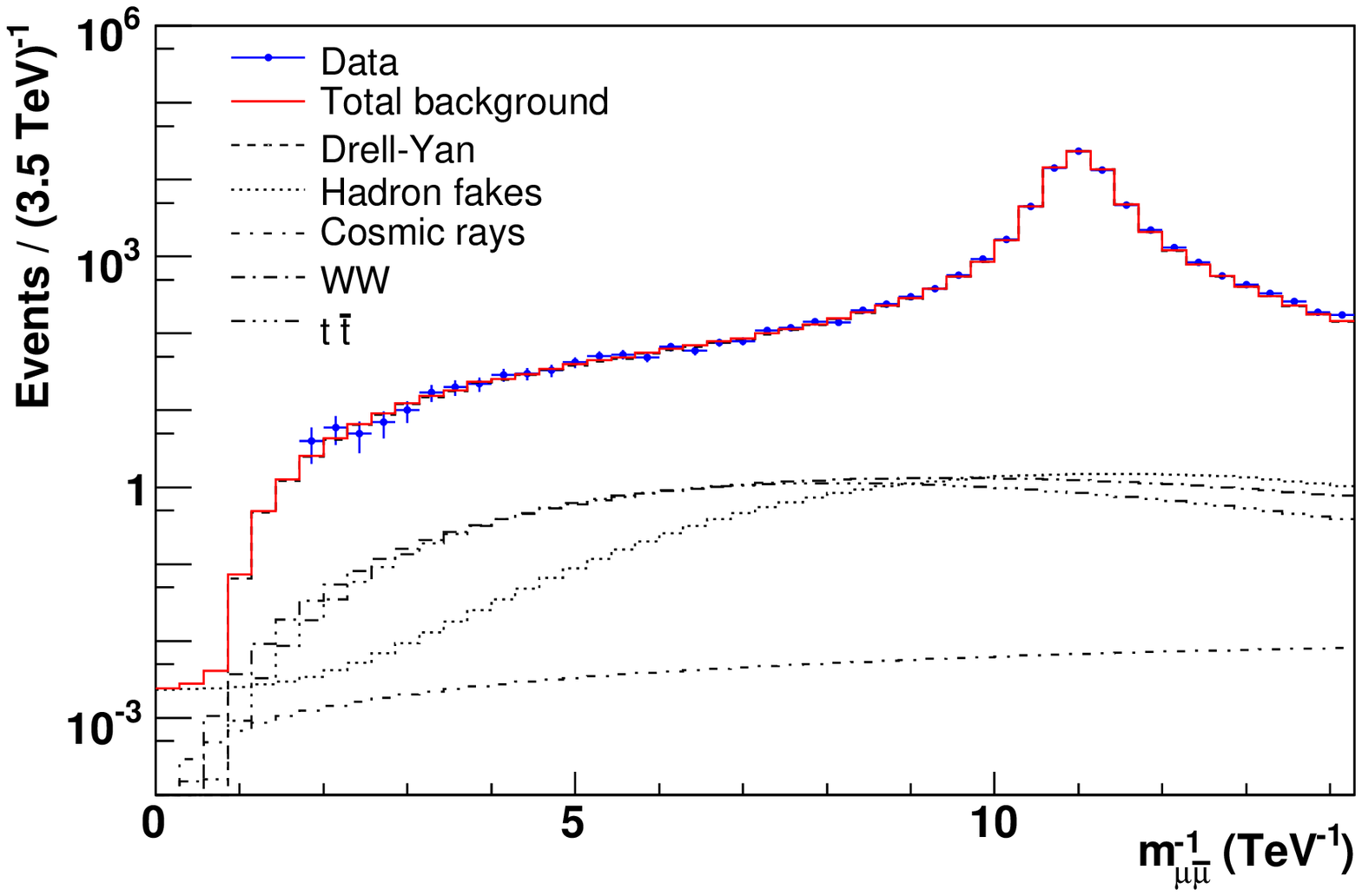}\put(-40,85){{ (b)}}\\
  \end{center}
  \vspace{-1pc}
  \caption{Invariant mass of electron pairs in D\O\ data compared to the SM expectation (a). Distribution of the di-muon inverse mass in CDF data (b).}
  \label{fig:dielectrons_dimuons}
\end{figure}

At the Tevatron, resonances from new particles have been searched for in invariant mass distributions of various di-particles combinations.
A peak in the di-lepton mass distribution could for instance sign the presence of an extra $Z'$ boson.
Recent measurements of the di-electron~\cite{D0_Zprime_ee} and di-muon~\cite{Aaltonen:2008ah} invariant mass distributions performed by the D\O\ and CDF Collaborations are presented in figure~\ref{fig:dielectrons_dimuons}.
An excess in the $e^+e^-$ invariant mass distribution at $\sim 240$~GeV was reported by the CDF Collaboration with a significance of $2.5\sigma$~\cite{Aaltonen:2008vx}. No such excess is observed in the recent analysis from D\O\ which uses a larger data sample of $3.6$~fb$^{-1}$~\cite{D0_Zprime_ee}.
A good agreement between data and the SM expectation is also seen in the di-muon mass spectrum measured by CDF. 
It is presented in terms of the variable $1/m_{\mu \mu}$ to better fit to the resolution of the trackers which is Gaussian in inverse momentum. 
Assuming a $Z'$ boson having the same couplings to SM fermions as the $Z$ boson, a lower limit of $1.03$~TeV was set on the $Z'$ mass.  
From the di-electron mass spectrum, $Z'$ bosons masses below $950$~GeV are excluded by D\O.

The invariant mass distribution of di-jets has also been investigated by the CDF Collaboration~\cite{Aaltonen:2008dn}. No deviation from NLO pQCD predictions was observed, allowing to exclude, for instance, the presence of $W'$ or $Z'$ bosons with masses below $840$ and $740$~GeV, respectively.

\begin{figure}[htbp]
  \begin{center}
    \includegraphics[width=0.24\textwidth]{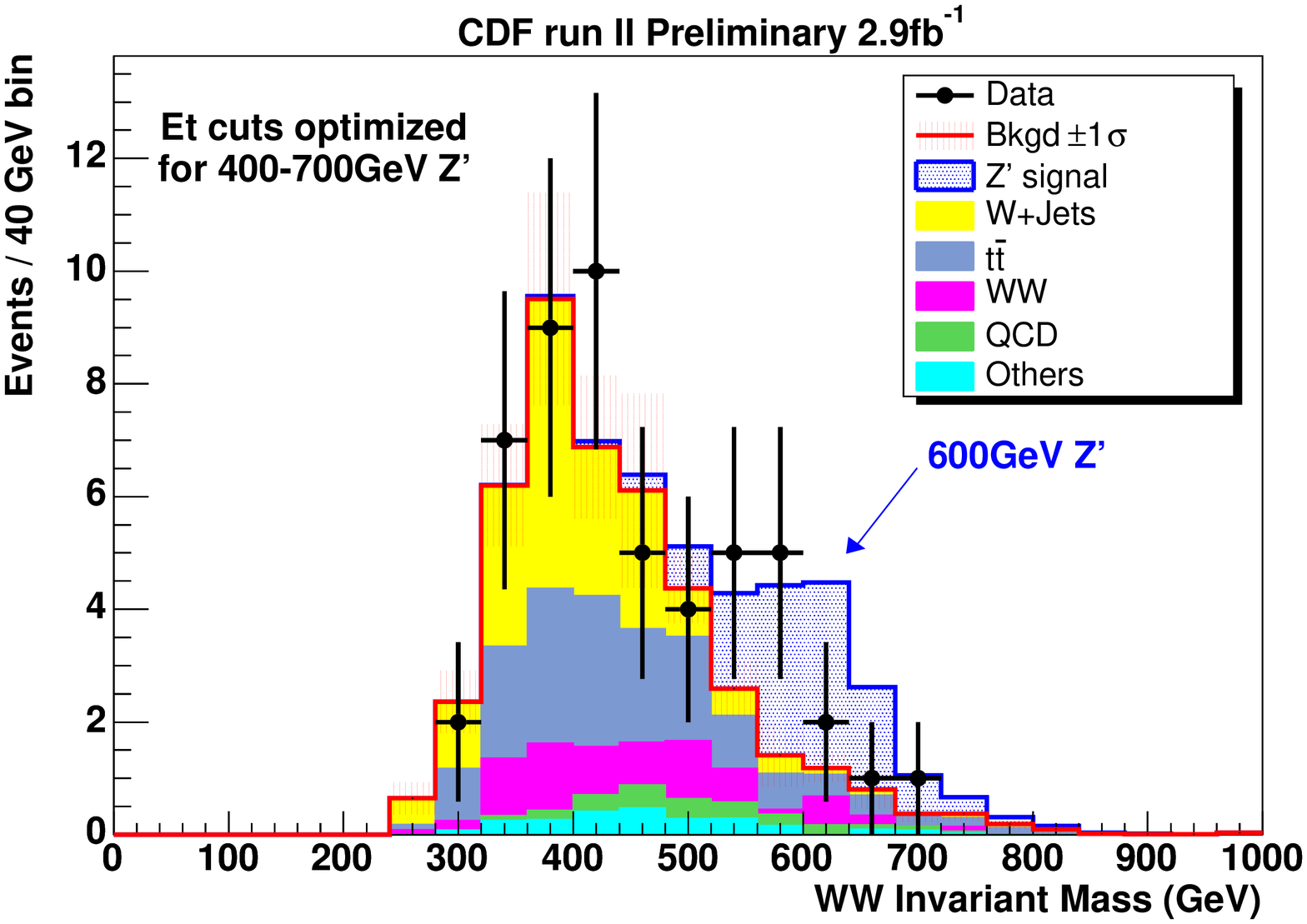}\put(-105,50){{ (a)}}
    \includegraphics[width=0.24\textwidth]{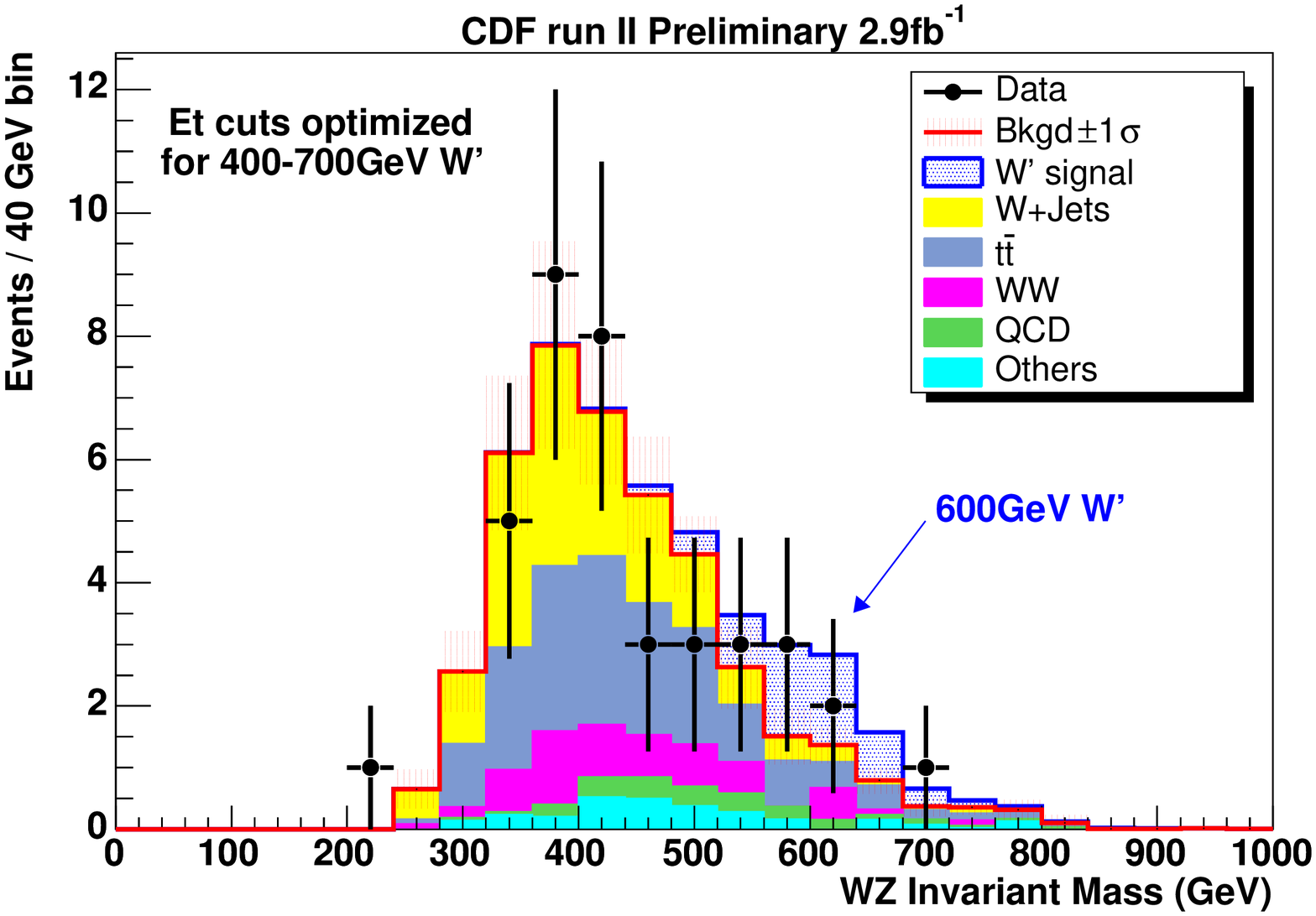}\put(-105,50){{ (b)}}
  \end{center}
  \vspace{-1pc}
  \caption{Di-boson invariant mass distributions for $WW$ (a) and $WZ$ (b) resonance searches, respectively, measured by CDF and compared to SM background expectations. Expected $W'$ and $Z'$ signals with a mass of $600$~GeV are added on top of the background in (a) and (b), respectively.}
  \label{fig:WW_WZ_resonances}
\end{figure}

A search for resonances decaying into a pair of gauge bosons, $W^+W^-$ or $W^\pm Z^0$, has been recently performed by the CDF experiment in final states where one $W$ boson decays into an electron and the other boson into two jets. 
Invariant mass distributions of the reconstructed boson pair are presented in figure~\ref{fig:WW_WZ_resonances}, when the reconstructed invariant mass of the two jets is constrained to be around the nominal $W$ or $Z$ boson masses, respectively. 
No significant excess over the SM prediction is observed.

\subsection{Isolated Lepton Events at HERA}

The production of $W$ bosons in $ep$ collisions at HERA has a cross-section of about $1$~pb. The leptonic decay of the $W$ leads to events with an isolated high transverse momentum lepton (electron, muon or tau) and missing total transverse momentum. Of particular interest are events with a hadronic system of large transverse momentum ($P_T^X$). A larger than expected rate of high $P_T^X$ events is observed by the H1 experiment~\cite{IL_H1} in the electron and muon channels. 
In the analysis of all available data, which amounts to a total luminosity of $478$~pb$^{-1}$, $18$ events are observed at $P_T^X > 25$~GeV for a SM expectation of $13.6 \pm 2.2$. Amongst them only $1$ event is observed in $e^-p$ collisions, compared to a SM expectation of $5.58 \pm 0.91$, while $17$ events are observed in the $e^+p$ data for an expectation of $8.0 \pm 1.3$.

\begin{figure}[htbp]
  \begin{center}
  \includegraphics[width=0.3\textwidth]{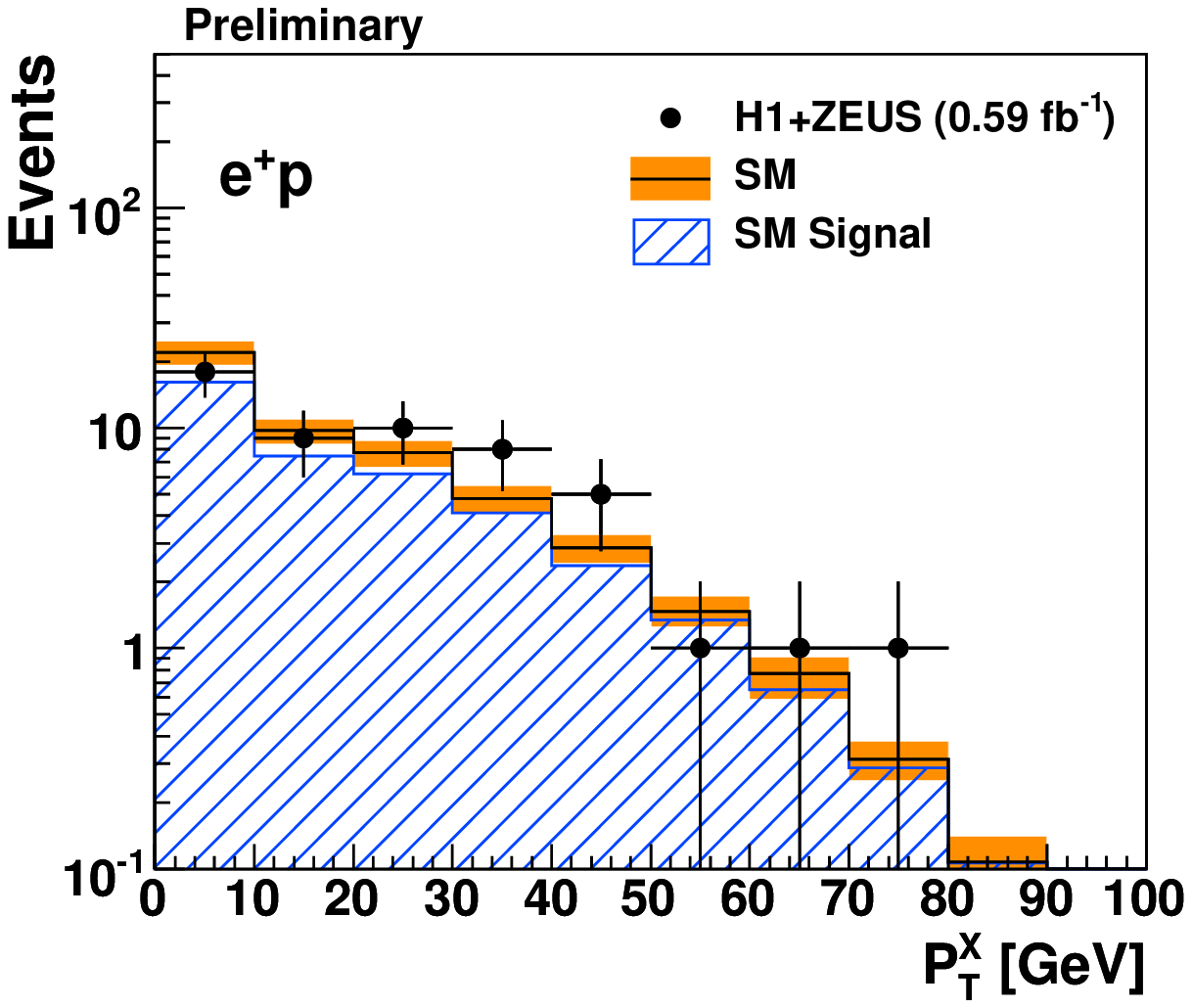}\put(-35,75){{ (a)}}\\
  \includegraphics[width=0.3\textwidth]{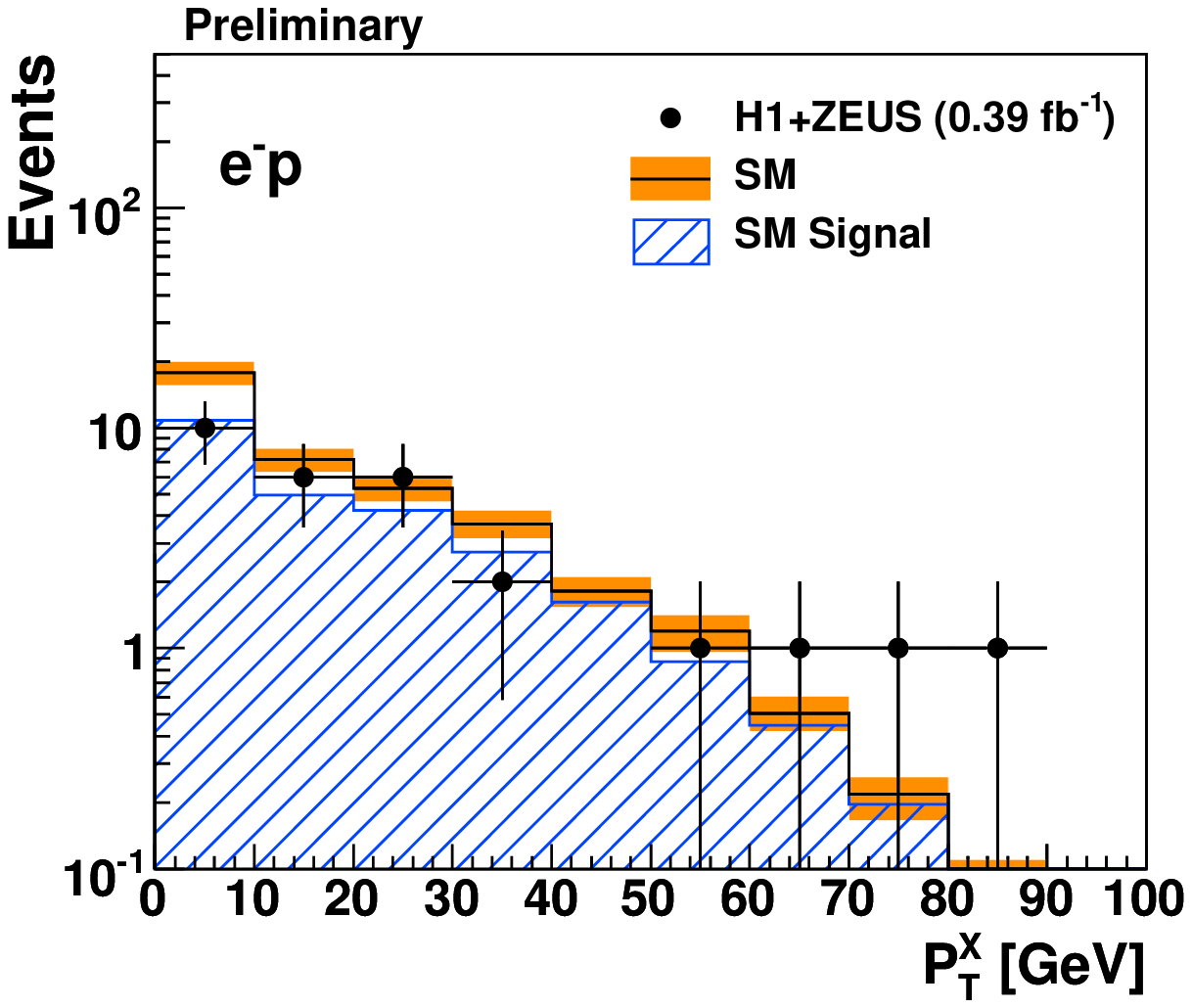}\put(-35,75){{ (b)}}
  \end{center}
  \vspace{-1pc}
  \caption{Hadronic transverse momentum distribution of isolated lepton events observed by H1 and ZEUS in $e^+p$ (a) and $e^-p$ (b) data samples. The total SM expectation is represented by the open histograms and the contribution from $W$ production by the hatched histograms.}
  \label{fig:isollep}
\end{figure}

The data samples of the H1 and ZEUS experiments have been used for a combined analysis performed in a common phase space~\cite{H1_ZEUS_IL}. The combined data set corresponds to a total integrated luminosity of $0.98$~fb$^{-1}$.
A total of $81$ events containing an isolated electron or muon and missing transverse momentum are observed in the data, compared to a SM expectation of $87.8 \pm 10.5$. At  $P_T^X > 25$~GeV, a total of $29$ events are observed compared to a SM prediction of $24.0 \pm 3.2$. In this kinematic region, $23$ events are observed in the $e^+p$ data compared to a SM prediction of $14.0 \pm 1.9$. Seventeen of the $23$ data events are observed in H1 data.
The observations in the $e^+p$ and $e^-p$ data sets are presented in figure~\ref{fig:isollep} where the $P_T^X$ distributions of both data sets are displayed.

\subsection{Multi-Lepton Production at HERA}

\begin{figure}[htbp]
  \begin{center}
    \includegraphics[width=0.3\textwidth]{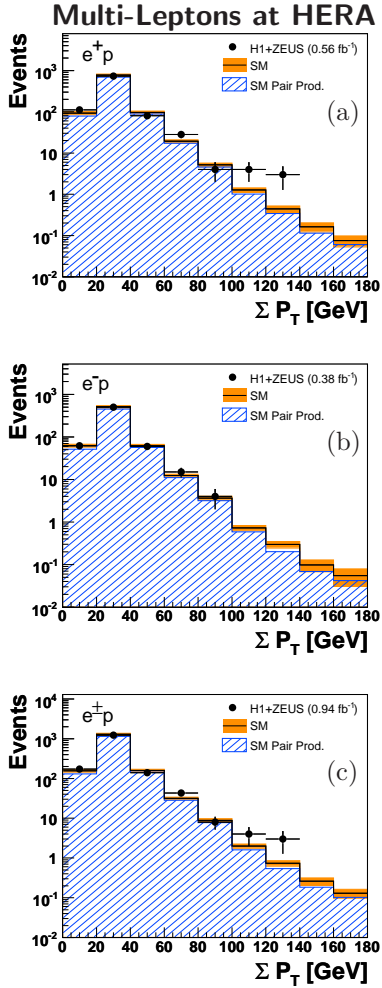}\put(-30,330){{ (a)}}\put(-30,205){{ (b)}}\put(-30,80){{ (c)}}
  \end{center}
  \vspace{-1pc}
  \caption{Distribution of the scalar sum of the lepton transverse momenta for multi-lepton events recorded by H1 and ZEUS compared to SM expectations, for $e^+p$ (a), $e^-p$ (b) and $e^\pm p$ (b) collisions.}
  \label{fig:ML_H1ZEUS}
\end{figure}

The main production mechanism for multi-lepton events at HERA is photon-photon collisions. All event topologies with high transverse momentum electrons and muons have been investigated by the H1 and ZEUS experiments~\cite{ML_H1ZEUS} using a total luminosity of $0.94$~fb$^{-1}$.
The measured yields of di-lepton and tri-lepton events are in good agreement with the SM prediction, except in the tail of the distribution of the scalar sum of transverse momenta of the leptons, $\sum P_T$ (see figure~\ref{fig:ML_H1ZEUS}). 
In $e^+p$ collisions, $7$  data events with at least two high $P_T$ leptons are observed with $\sum P_T > 100$~GeV compared to a SM prediction of $1.94 \pm 0.17$, corresponding to a probability of $0.4$\%. No such events are observed in $e^-p$ collisions for a similar SM expectation of $1.19 \pm 0.12$.

\subsection{Signature Based Searches with Photons}

The CDF experiment has developed a series of model-independent signature based searches, looking for deviations from the SM in various final state topologies containing a photon.
Such analyses are based on the ability to identify and define a priori objects which can be charged leptons, jets or heavy flavour quarks, electroweak gauge bosons, or non-interacting particles identified by \mbox{\ensuremath{\slash\kern-.7emE_{T}}}, independently of the considered final state.

Different event topologies with two photons and a third object ($e$, $\mu$, $\tau$, $\gamma$ or \mbox{\ensuremath{\slash\kern-.7emE_{T}}}) have been investigated by CDF in a data sample of $2$~fb$^{-1}$ of integrated luminosity~\cite{CDF_gammaX}. 
No deviations from the SM expectations were observed in these final states.

Recently, $\gamma b j$\mbox{\ensuremath{\slash\kern-.7emE_{T}}}~\cite{Aaltonen:2009dn} and $\ell b j$\mbox{\ensuremath{\slash\kern-.7emE_{T}}}~\cite{Aaltonen:2009iw} topologies have also been investigated by CDF.
Numbers of data events observed in each channel were found to be in good agreement with SM expectations.

\subsection{General Searches}

The extension of signature based searches are global model independent searches, extending to all possible topologies at high $P_T$.
In this approach all events are classified into exclusive event classes according to the number and types of objects detected in the final state.
A comparison is then performed between the numbers of events in each class and the expectations from all SM processes.

\begin{figure}[htbp]
  \begin{center}
    \includegraphics[width=0.484\textwidth,angle=-90]{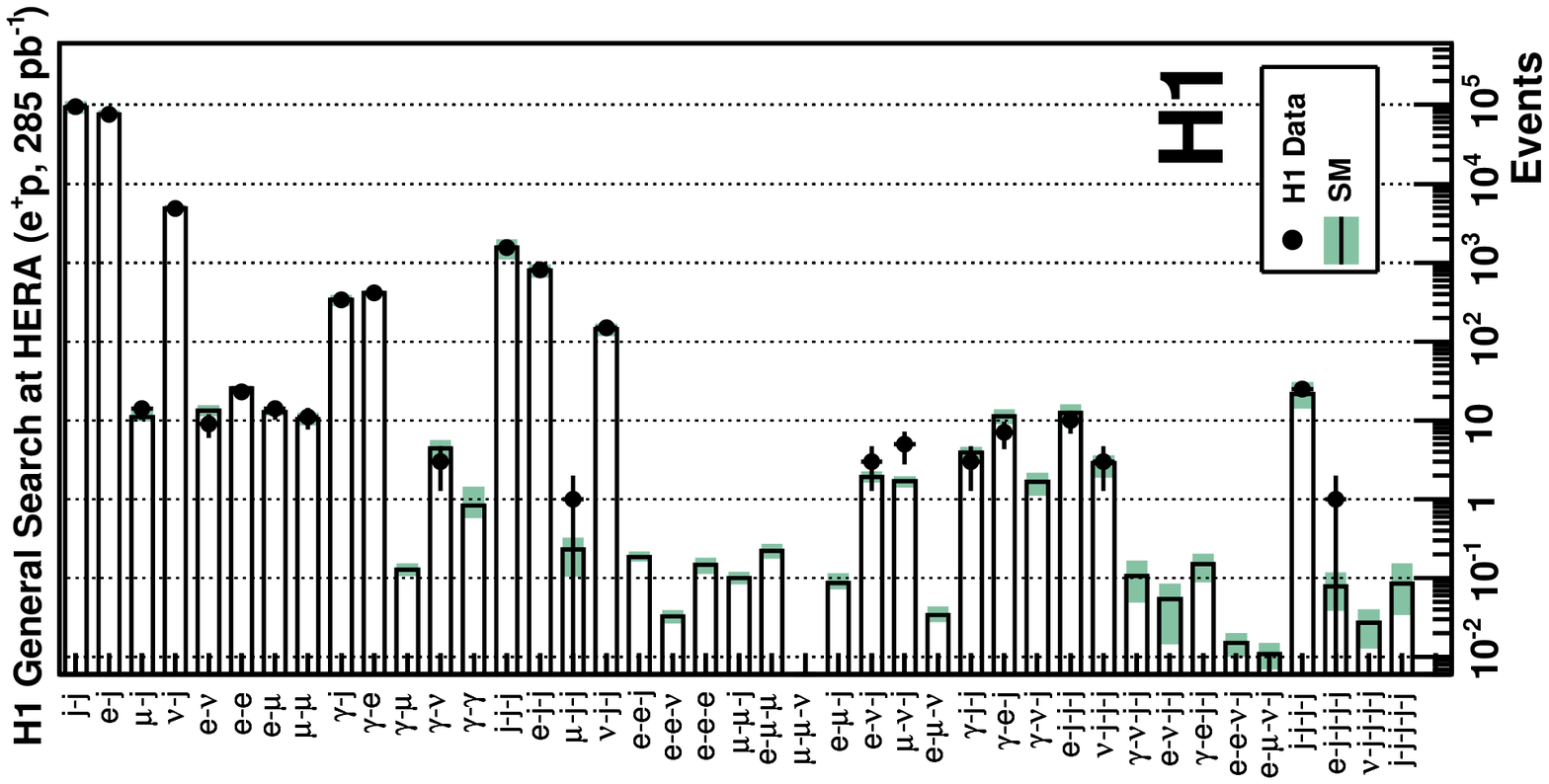}\put(-17,-160){{(a)}}
    \includegraphics[width=0.484\textwidth,angle=-90]{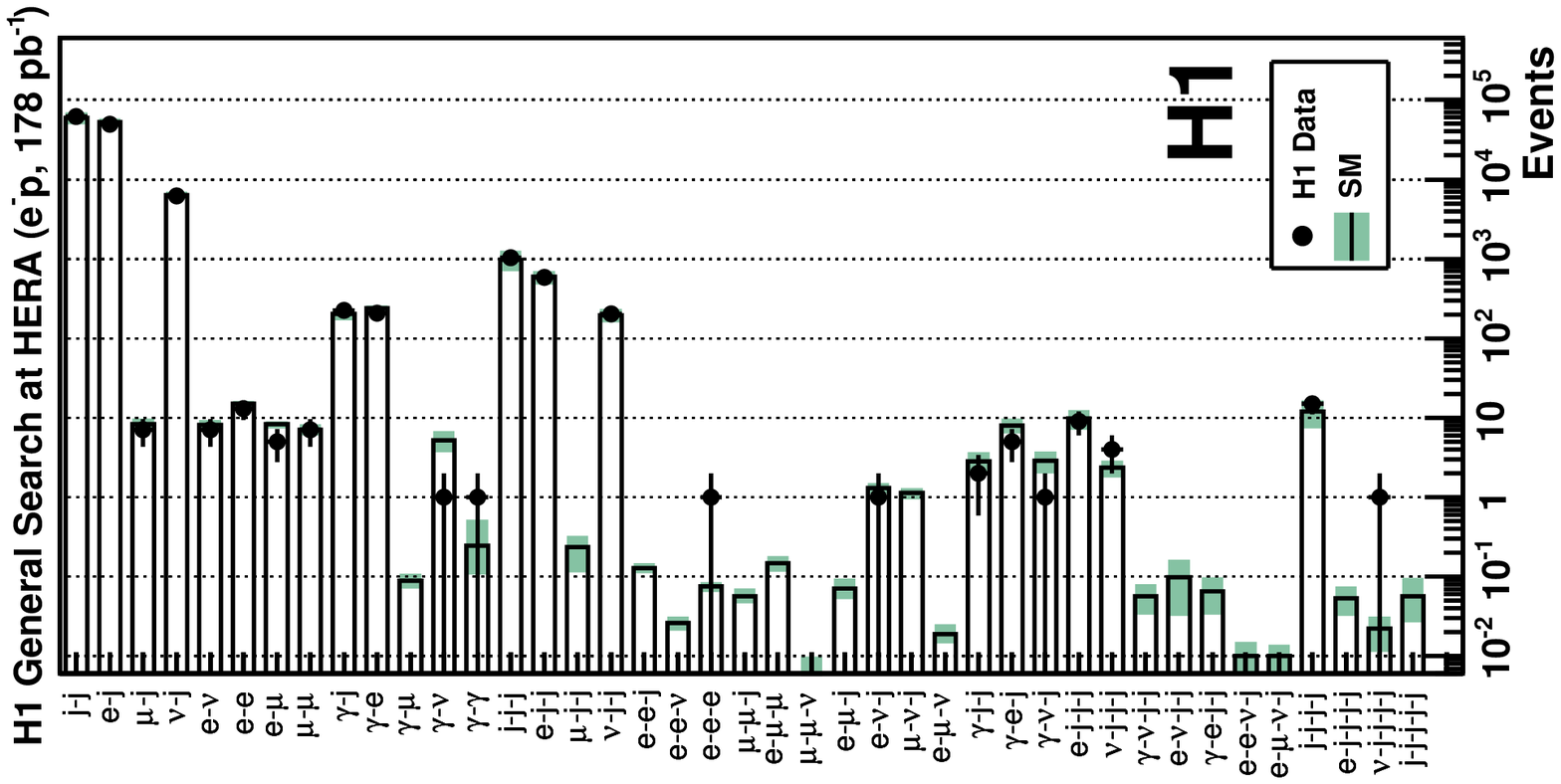}\put(-17,-160){{(b)}}
  \end{center}
  \vspace{-1pc}
  \caption{The data and the SM expectation in event classes investigated by the H1 general search. Only channels with observed data events or a SM expectation greater than one event are displayed. The results are presented separately for $e^+p$ (a) and $e^-p$ (b) collision modes.}
  \label{fig:H1_GS}
\end{figure}

Such a broad range signature based search has been pioneered by the H1 Collaboration using data from the first running phase of HERA, which were mainly from $e^+p$ collisions~\cite{GS_H1_hera1}.
It has been recently updated to the full data set of H1, which amount to an integrated luminosity of $463$~pb$^{-1}$ and includes $e^-p$ collision data~\cite{GS_H1}.
All final states containing at least two objects ($e$, $\mu$, $j$, $\gamma$, $\nu$) with 
$P_T >$~$20$~GeV in the polar angle range  $10^\circ < \theta < 140^\circ$ are investigated.
The observed and predicted event yields in each channel are presented in figure~\ref{fig:H1_GS}(a) and (b) for $e^+p$ and $e^-p$ collisions, respectively.
The good agreement observed between data and SM prediction demonstrates the good understanding of the detector and of the contributions of the SM backgrounds.

In each channel, a systematic scan of distributions of the invariant mass, $M_{all}$, and of the scalar sum of transverse momenta, $\sum P_T$ of all identified objects has been performed to look for regions of largest deviations to the SM. A statistical analysis is then used to quantify the significance of the observed deviations. The largest deviation is found in $e^+p$ data, in the $e$-$e$ channel which corresponds to the topology of multi-lepton events. It corresponds to a probability of 0.0035. The probability to observe
a SM fluctuation with that significance or higher for at least one event class is 12\%.
In addition, the final state topologies are also evaluated in terms of angular distributions and energy sharing between final state particles, with a good agreement observed between data and the SM expectation.

A similar approach is followed at the Tevatron by the CDF and D\O\ experiments.
The CDF Collaboration considered $399$ final states containing electrons, muons, taus, photons, jets, $b$-jets and \mbox{\ensuremath{\slash\kern-.7emE_{T}}}\ in a data sample of $2$~fb$^{-1}$~\cite{Aaltonen:2008vt}. 
The shapes of $19650$ distributions have been studied using a dedicated algorithm, called {\sc VISTA}~\cite{Aaltonen:2007dg}, and discrepancies between data and the SM expectation have been found in $555$ of them.
The origin of these discrepancies is however attributed to an inadequate modeling of soft QCD effects in the simulation, rather than to signs of new physics.
In a second step, high $P_T$ tails of the distributions were investigated using the {\sc SLEUTH} algorithm~\cite{Aaltonen:2007dg}. No excess beyond what is expected from statistics was observed.
Using a similar framework based on {\sc VISTA} and {\sc SLEUTH} algorithms, the D\O\ experiment investigated $180$ final states containing at least one lepton, using $\sim 1$~fb$^{-1}$ of data~\cite{D0_GS}.
Only four out of the $180$ final states present a statistically significant discrepancy between data and SM expectations. However, the observed deviations are attributed to difficulties in modelling the SM background or the detector response in the given final states and do not point to new physics.

\subsection{Conclusions}

Despite intensive searches performed in data of high energy colliders, no convincing sign of physics beyond the SM has been found so far.

Two years after the end of HERA running, the H1 and ZEUS Collaborations are delivering their final results based on an integrated luminosity of $1$~fb$^{-1}$, combining their data in common analyses. 
The most significant deviation to the SM expectation observed in HERA data concerns intriguing  multi-lepton events in $e^+ p$ data.

At the Tevatron, many different models of new physics are tested with ever increasing sensitivity.
Complementing to these model-dependent searches, more general tests of the SM validity at the highest energies are also being developed through signature based searches. 
But, so far, no discoveries are to be reported and stringent constraints on models are set.

Most likely, new physics will remain hidden until the advent of the LHC era.



\end{document}